# Erbium-implanted tellurite waveguides with low-temperature post-implantation activation and signal enhancement

**Yuxuan Gao[1], Batoul Hashemi[1,*], Bruno L. Segat Frare[1], Niloofar Majidian Taleghani[1], Pooya Torab Ahmadi[1], Henry C. Frankis[1], Ponnambalam Ravi Selvaganapathy[2], Jonathan D. B. Bradley[1], Peter Mascher[1], and Andrew P. Knights[1]**

[1] Department of Engineering Physics, McMaster University, 1280 Main Street West, Hamilton, ON L8S 4L7, Canada
[2] Department of Mechanical Engineering and the School of Biomedical Engineering, McMaster University, 1280 Main Street West, Hamilton, Ontario L8S 4L7, Canada
*Author to whom any correspondence should be addressed.

**E-mail :** hashemb@mcmaster.ca

## Abstract

In this paper, we demonstrate erbium ion implantation and signal enhancement in tellurium oxide hybrid waveguides. Silicon nitride strips with a width of 2 µm and a height of 100 nm were clad with a 110-nm-thick tellurium oxide layer to form hybrid waveguides, followed by erbium ion implantation at an energy of 200 keV and a dose of $1\times10^{15}$ ions/cm$^2$, with a projected peak implantation depth of approximately 50 nm into the tellurium oxide layer. After low-temperature annealing at 150 °C for 30 minutes, the propagation loss decreased from 1.7 to 0.9 dB/cm, while the erbium lifetime increased from 40 µs to over 800 µs. We measure a small-signal enhancement of 9 dB in an 11-cm-long waveguide at a wavelength of 1550 nm. These results demonstrate progress towards a low-temperature post-ion implantation process for incorporating erbium and other rare earth ions into tellurium oxide films for integrated photonic applications.

## 1. Introduction

Erbium-doped fiber amplifiers (EDFAs) have played a central role in optical communication systems since their development in the 1980s, particularly for signal amplification in the C- and L-bands used in dense wavelength division multiplexing (DWDM) [1,2]. The increasing demand for compact and scalable photonic systems has motivated the development of erbium-doped waveguide amplifiers (EDWAs), which aim to provide optical gain directly on integrated photonic platforms[3–7]. To realize high-performance EDWAs, a range of host materials and fabrication approaches have been investigated. Suitable EDWA platforms require high Er solubility, low optical propagation loss, uniform film deposition over large areas, compatibility with CMOS processing, and practical deposition rates. Erbium-doped crystalline materials, such as lithium niobate [8–15], as well as glass and amorphous hosts, including aluminum oxide ($Al_2O_3$) [16–23], tantalum pentoxide ($Ta_2O_5$) [24–27], silicate glasses [28–31], and tellurium oxide ($TeO_2$) [32–35], have been explored for integrated optical gain.

Several techniques have been employed to incorporate Er ions into optical host materials, including co-sputtering, ion implantation, and co-doping strategies. While co-sputtering of tellurium and rare-earth elements in an oxygen ambient has previously enabled $TeO_2$-based EDWAs [32–35], ion implantation provides an alternative non-equilibrium approach that enables precise control over dopant depth and spatial distribution, while allowing high dopant concentrations beyond equilibrium solubility limits [4,36]. This approach has been widely investigated for Er doping of thin-film photonic materials, including silica, phosphosilicate, aluminum oxide, silicon nitride, and lithium niobate platforms [36–44]. Recently, Er implantation in ultralow-loss $Si_3N_4$ photonic integrated circuits has enabled high-performance on-chip amplification, with a reported output power of 145 mW and small-signal gain above 30 dB [40]. These results demonstrate the strong potential of implantation based Er doping for integrated photonics. Ion implantation also allows co-doping with additional ions, such as ytterbium or oxygen, to improve Er-related emission properties. Yb co-doping can enhance Er excitation through energy transfer from $Yb^{3+}$ to $Er^{3+}$, while oxygen co-implantation can promote the formation of optically active Er–O complexes and improve radiative efficiency [36,45]. Er implantation has also been reported to suppress undesired cooperative upconversion compared with co-sputtered films [4]. However, direct implantation into tightly confined $Si_3N_4$ waveguides can require high implantation energies and subsequent relatively high-temperature thermal annealing to recover implantation-induced damage and activate optically efficient Er centers [40–42,46,47]. Therefore, alternative host layers and lower-temperature integration routes remain of interest.

Tellurite is an attractive host material for Er-doped waveguide amplifiers because of its relatively high refractive index, broad emission bandwidth, high rare-earth solubility, reduced cross-relaxation between Er ions, and strong optical nonlinearity [32–35,48]. $TeO_2$ thin films can also be deposited using wafer-scale, low-temperature processes, making them well suited for integration with a range of silicon photonics platforms, including silicon-on-insulator (SOI) and $Si_3N_4$ [34,35,49–51]. This hybrid approach combines the favorable rare-earth-hosting properties of tellurium oxide with the mature fabrication, compact device geometries, and passive waveguiding capabilities of silicon photonic platforms[34,49,52–54]. $TeO_2$-based active layers have been integrated with both SOI and $Si_3N_4$ platforms to realize compact lasers and optical amplifiers [34,51]. Such heterogeneous integration also provides a promising route toward incorporating active and nonlinear functionalities into silicon-based photonic circuits [35,55]. Ion implantation has previously been investigated in tellurite-based glasses for waveguide formation or modification of rare-earth-doped glass properties, including $H^+$, $O^+$, and ion-beam-irradiated $Er^{3+}$- or $Yb^{3+}$-doped tellurite glasses [56–58]. However, direct implantation of erbium ions into $TeO_2$ thin films remains largely unexplored. Specifically, the impact of optical doping via ion implantation on the degradation of tellurite films, and the recovery of the films using annealing is not well-known.

In this work, we demonstrate $Er^+$ implantation into a $TeO_2$ layer followed by low-temperature post-implantation annealing. Low-temperature annealing is shown to activate the implanted Er ions while reducing waveguide background loss through partial recovery of implantation-induced optical damage. The fabricated hybrid $TeO_2$-$Si_3N_4$ waveguide spirals exhibit an $Er^{3+}$ luminescence lifetime exceeding 800 μs and a signal enhancement of approximately 9 dB, showing a pathway to optical gain rare-earth-ion implanted tellurite on-chip devices.

## 2. Design and fabrication

Figure 1(a) shows the cross-section of the EDWA structure studied in this work. The hybrid $TeO_2$–$Si_3N_4$ platform combines the rare-earth-hosting properties of $TeO_2$ with the low-loss guiding characteristics of $Si_3N_4$. The waveguide geometry allows the optical mode to overlap with the Er-implanted $TeO_2$ layer, while the implantation profile and associated damage remain largely confined to the $TeO_2$ rather than the $Si_3N_4$ core. The uncladded $Si_3N_4$ waveguide, with a height of 100 nm and width of 2 μm, was fabricated by LioniX International using stepper lithography and reactive ion etching[59]. The low-confinement $Si_3N_4$ geometry allows the fundamental TE mode to extend into the cladding, increasing the overlap between the optical field and the active Er ions. The spiral waveguide has a total length of 11 cm and includes 250-μm-long linearly tapered edge couplers on both sides. A 110-nm-thick $TeO_2$ film was deposited on the $Si_3N_4$ chip by room-temperature radio-frequency reactive sputtering using a Lesker PVD Pro 200 system at the Centre for Emerging Device Technologies (CEDT), McMaster University, using a process similar to that described in [34,50]. Erbium ions were introduced into the $TeO_2$ film by ion implantation with an energy of 200 keV and a dose of $1\times10^{15}$ ions/cm$^2$, carried out at the UK Ion Beam Centre at the University of Surrey.

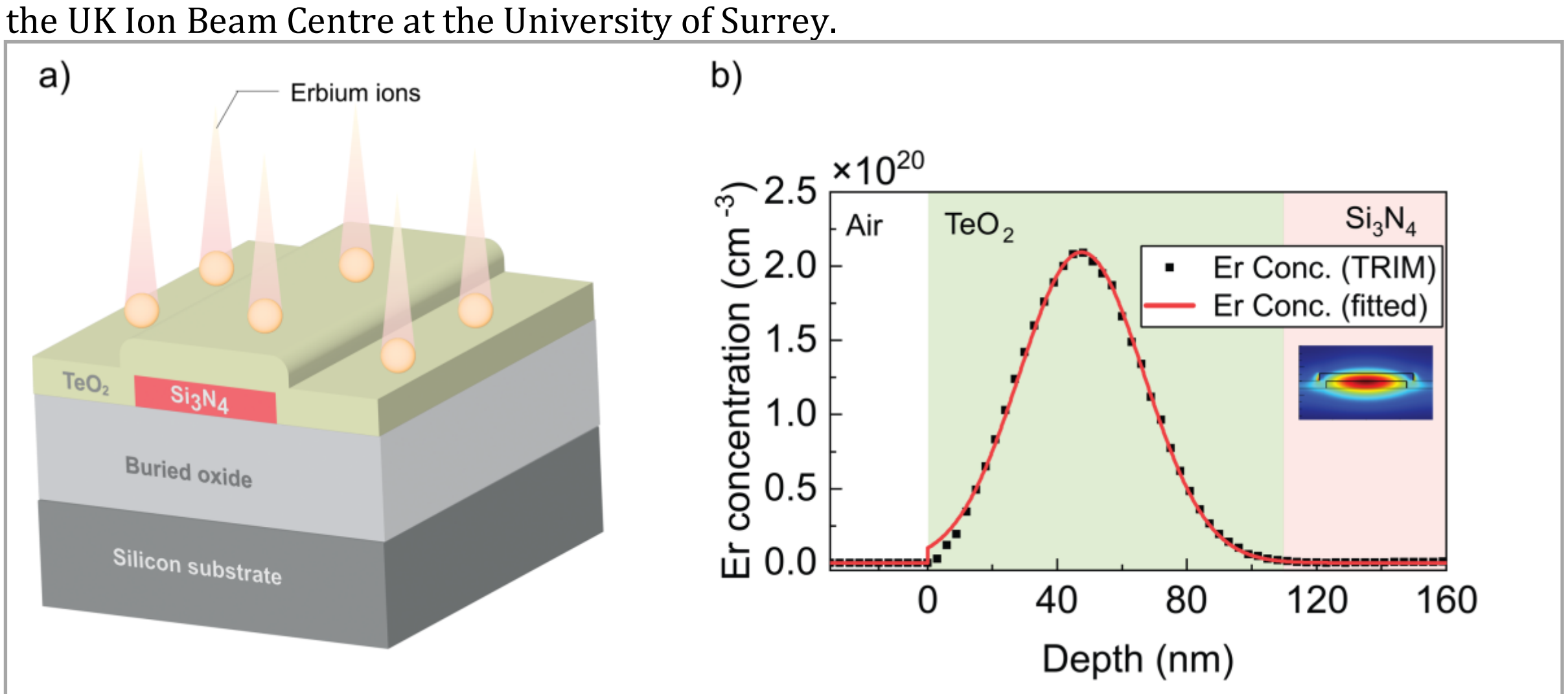

**Figure 1.** a) Diagram of the erbium-implanted $TeO_2$-coated $Si_3N_4$ waveguide structure. b) Calculated (black dots) and fitted (red line) erbium concentration using TRIM simulations, and the simulated optical transverse electric mode intensity (inset).

The Er concentration profile, plotted in Fig. 1(b), was calculated using Transport of Ions in Matter (TRIM) simulations for Er implanted into $TeO_2$, giving a peak concentration of $2\times10^{20}$ cm$^{-3}$ at a depth of approximately 50 nm from the surface of the $TeO_2$ film. The $TeO_2$ layer thickness was chosen to increase the overlap between the guided mode and the implanted Er distribution, considering the short ion projected range of approximately 50 nm, as shown in the inset of Fig. 1(b), while maintaining low propagation loss. This configuration enables efficient excitation of $Er^{3+}$ ions residing in the $TeO_2$ film. Although a deeper implant profile, multiple implantation steps, and/or a thicker $TeO_2$ layer could further increase the modal overlap by better aligning the Er distribution with the guided optical mode, substantially higher implantation energies would be required in all of those cases. Moreover, only a negligible amount of Er penetrates beyond the $TeO_2$ layer, leaving the low-loss $Si_3N_4$ waveguide largely unaffected for the purposes of this study. Following ion implantation, the sample underwent a low temperature anneal at 150 °C for 30 min in nitrogen and was subsequently characterized. To investigate whether additional defect recovery could be achieved while maintaining a low thermal budget, the same sample was then annealed at 180 °C for a further 30 min in nitrogen.

## 3. Experimental characterization

Gain measurements were conducted using a double-sided pumping arrangement, as shown in Fig. 2. At the input side, a tunable C-band signal laser operating from 1530 to 1565 nm was combined with a 1470 nm pump laser diode using a wavelength-division multiplexer (WDM). The combined light was coupled into the device using a tapered fiber with a 2.5 μm spot size. At the output side, a second 1470 nm pump laser diode was coupled into the device through the collection fiber to provide counter-propagating pump light. Polarization paddles were adjusted to launch the fundamental transverse-electric (TE) mode. Light exiting the waveguide was collected using a second tapered fiber. The outcoupled signal then passed through a WDM and a 1500 nm edge-pass filter to remove residual pump light before being detected by a photodetector. Lifetime measurements were performed using a function generator and an electrical oscilloscope. A 100 Hz periodic square wave with a 50% duty cycle was used to modulate the pump laser, and the spontaneous-emission decay was recorded on the oscilloscope. Single-bus microring resonators with a waveguide width of 1.2 μm and a radius of 500 μm were characterized to independently evaluate implantation- and annealing-induced changes in intrinsic propagation loss. The microring measurements were therefore used primarily to assess the relative evolution of waveguide loss across the different processing stages [49].

## 4. Results and discussion

Figure 3 shows representative ring-resonator transmission spectra measured before implantation (yellow), after Er implantation (orange), and after annealing at 150 °C (blue). The ring resonators were undercoupled, with a bus-to-ring gap of 2.8 μm, and representative resonances outside the main Er absorption band were selected to extract the loaded and intrinsic quality factors. For the 1.2 μm-wide ring resonator with a radius of 500 μm, the loaded quality factors before implantation, after implantation, and after annealing were $4.7 \times 10^5$, $1.6 \times 10^5$, and $2.8 \times 10^5$ respectively. The corresponding intrinsic quality factors were $5.6 \times 10^5$, $1.8 \times 10^5$, and $3.3 \times 10^5$ at resonance wavelengths of 1570.09, 1580.10, and 1581.04 nm, respectively. These intrinsic quality factors correspond to estimated propagation losses of approximately 0.5, 1.7, and 0.9 dB/cm, demonstrating a substantial increase in optical loss following Er implantation and a partial recovery after low-temperature annealing [49,60]. The increased waveguide propagation loss (from approximately 0.5 to 1.7 dB/cm) is attributed to implantation-induced defects in the $TeO_2$ cladding. Low-temperature annealing at 150 °C was employed to partially repair the implantation-induced defects and improve waveguide transmission, reducing the propagation loss to approximately 0.9 dB/cm. Annealing at 180 °C was also investigated but provided no appreciable additional improvement compared with the 150 °C treatment. Although higher temperature annealing might further reduce the implantation-induced loss, it can also promote crystallization of the $TeO_2$ film, resulting in increased scattering and propagation loss [61–64]. Furthermore, the 150 °C annealing temperature is substantially lower than the temperatures of approximately 1000 °C previously employed in other ion-implanted waveguide systems [40], and is compatible with Back-End-of-Line (BEOL) processing. This post-implantation annealing process, combined with a limited ion implantation energy of 200keV points to a fabrication process which is compatible also with the majority of implantation tools in service.

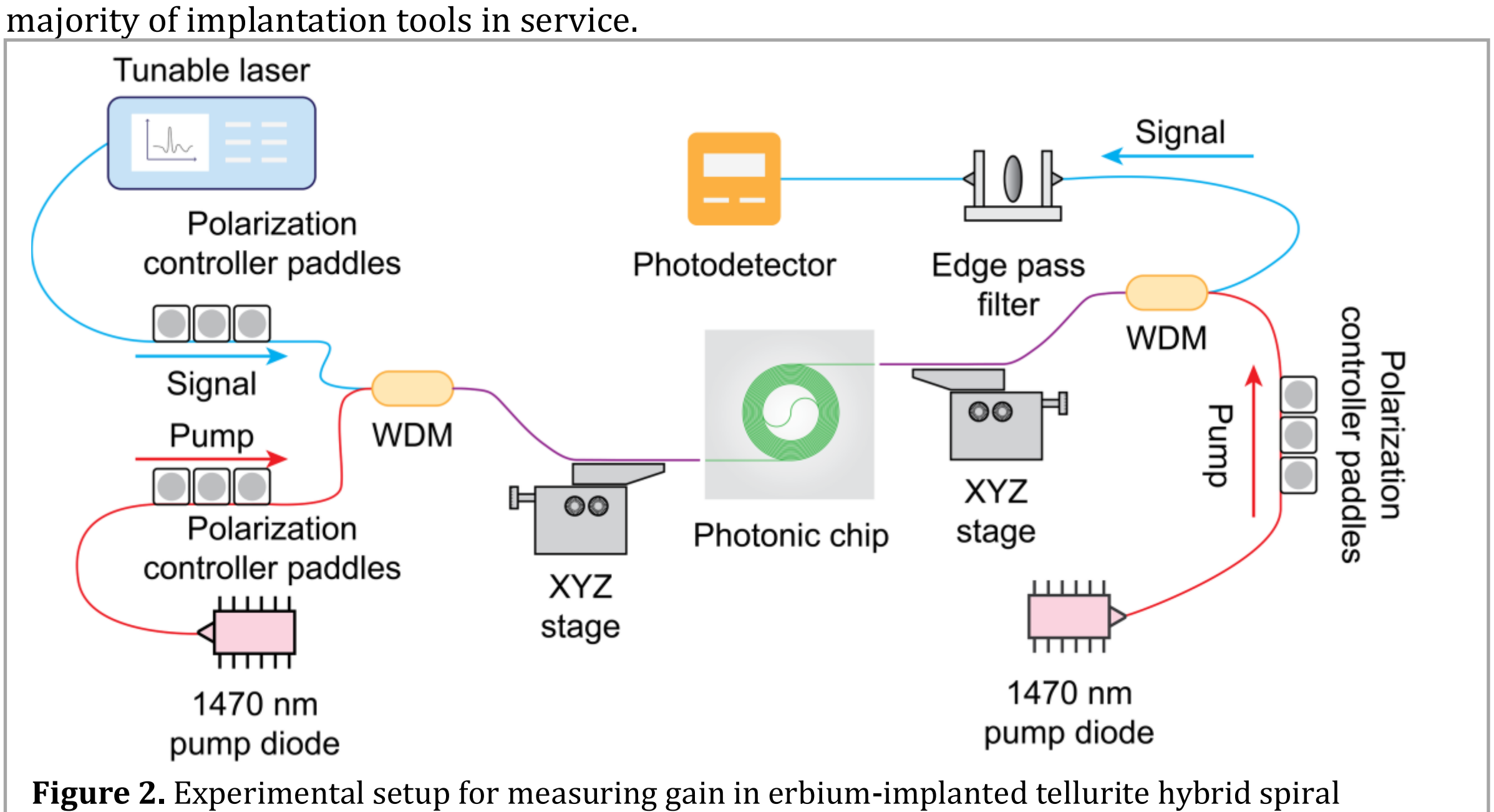

**Figure 2.** Experimental setup for measuring gain in erbium-implanted tellurite hybrid spiral waveguides.

In addition to the improved $TeO_2$ film quality, an increased $Er^{3+}$ photoluminescence lifetime was also observed after annealing. Fig. 4 demonstrates a 20-fold increase in lifetime to approximately 800 μs after annealing, as determined by exponential curve fitting, compared with approximately 40 μs before annealing. This increase is again attributed to the repair of defects that acted as non-radiative recombination centers [4]. Additionally, annealing can reduce hydroxyl (OH) groups in the host material, preventing photoluminescence quenching [37,65,66].

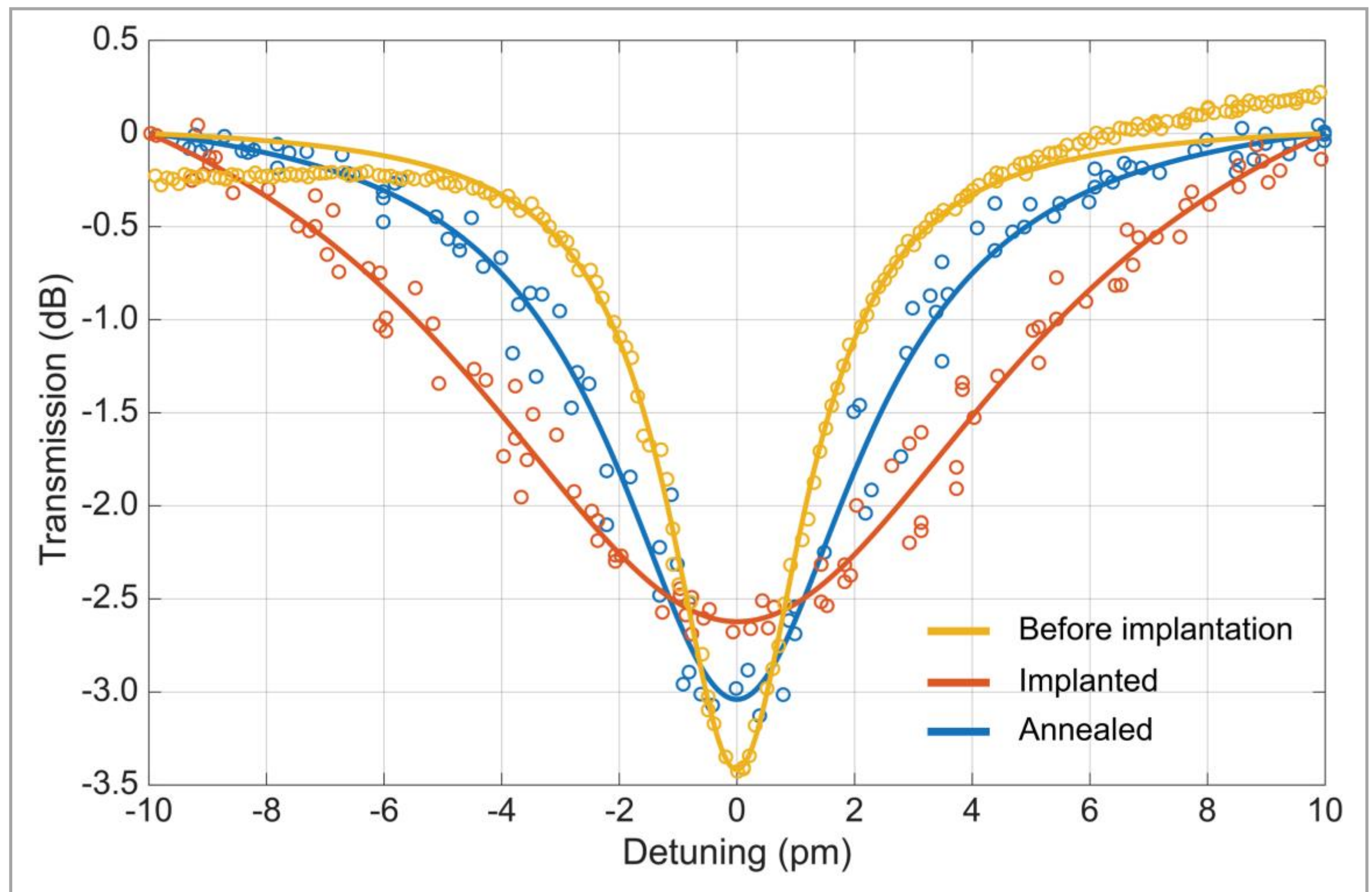

**Figure 3.** Transmission spectra and Lorentzian fittings indicating the Q-factors of the microrings before implantation, after implantation, and after annealing.

The passive transmission of the 11 cm spiral waveguide was measured at different fabrication stages: before implantation, after Er implantation, and after annealing. For the hybrid waveguide prior to Er implantation, the propagation loss over the 11 cm long spiral, excluding the estimated fiber-to-chip coupling loss of approximately 3 dB for both facets, was approximately 4.5 dB for the wavelength range of 1475–1640 nm. The additional optical loss induced by Er implantation was then estimated by comparing the transmission spectra measured before and after implantation, as illustrated in Fig. 5(a), assuming consistent fiber-to-chip edge coupling loss across all measurements.

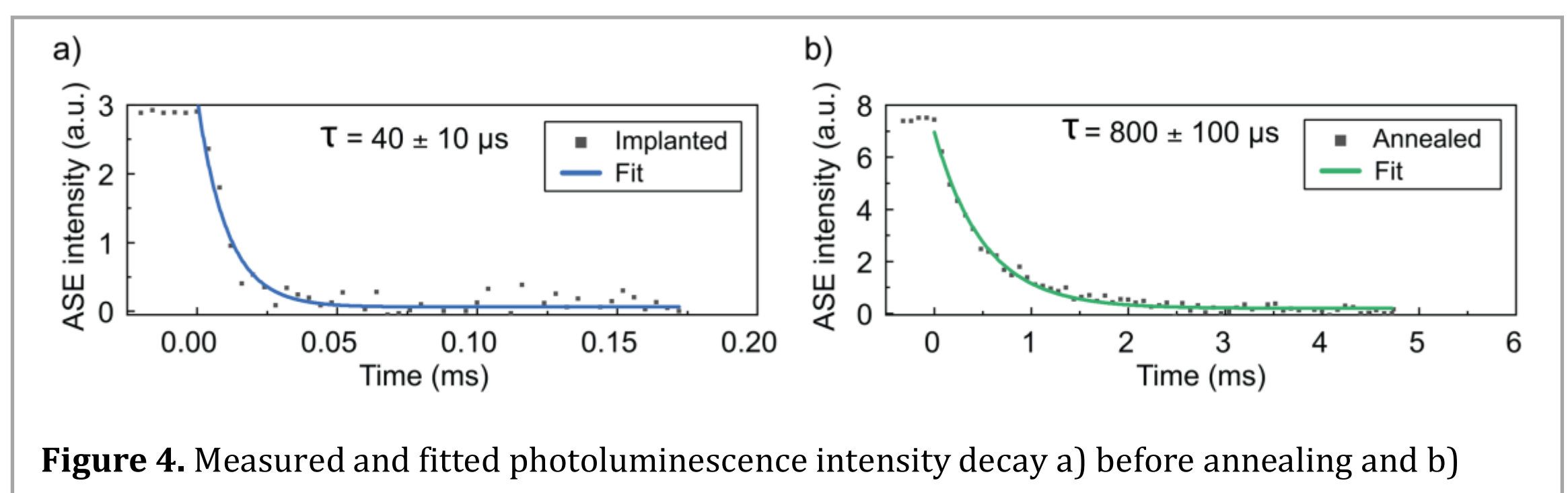

**Figure 4.** Measured and fitted photoluminescence intensity decay a) before annealing and b) after annealing.

After implantation, the measured loss near 1620 nm increased by approximately 8.8 dB (0.8 dB/cm), where Er absorption is negligible and the excess loss is therefore attributed primarily to implantation-induced background loss. At the Er absorption peak near 1530 nm, the total implantation-induced excess loss was approximately 14.3 dB (1.3 dB/cm), consisting of approximately 8.8 dB of background loss (0.8 dB/cm) and 5.5 dB of Er-related absorption (0.5 dB/cm). Following annealing at 150 °C in nitrogen for 30 minutes, the background-loss contribution decreased from approximately 8.8 to 5.5 dB (0.8 to 0.5 dB/cm), indicating partial recovery of implantation-induced defects [57,67]. In comparison, the Er-related absorption changed by less than 2.2 dB (0.2 dB/cm), suggesting that the low-temperature annealing caused only limited diffusion of the implanted Er ions.

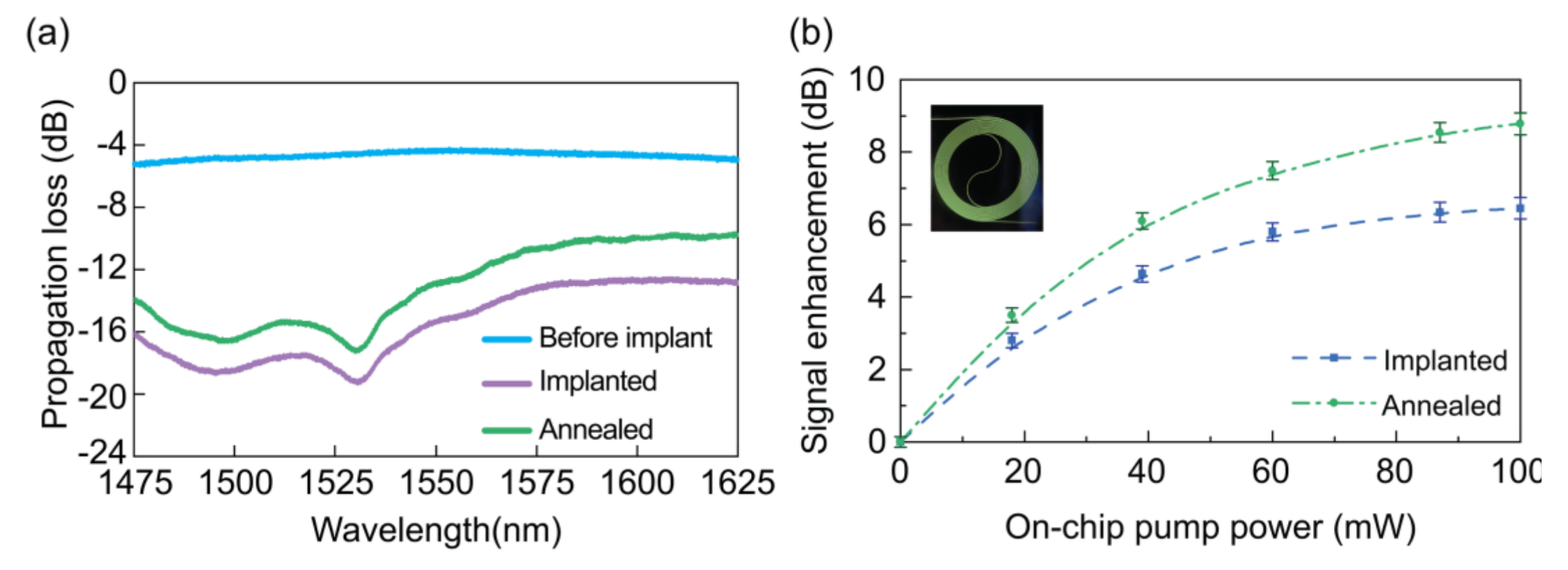

**Figure 5.** a) Measured propagation loss spectrum (excluding fiber chip edge coupling loss) of 11-cm-long waveguide before implantation(blue), after implantation (purple) and annealing (green); b) Measured signal enhancement factor (dB) at 1550 nm as a function of on-chip optical pump power (mW) at 1470 nm.

The signal enhancement of the waveguide after implantation and after annealing is shown in Fig. 5(b). The inset shows the strong green emission from second order $Er^{3+}$ up conversion observed after annealing. The signal enhancement was measured at 1550 nm as a function of pump power using a 1470 nm laser diode, while the on-chip signal power was maintained at approximately −20 dBm to avoid gain saturation. Peak signal enhancements of approximately 6.6 and 8.8 dB, corresponding to approximately 0.6 and 0.8 dB/cm were obtained from the 11 cm waveguide after implantation and after annealing, respectively. The net gain, calculated by subtracting the Er absorption loss and background propagation loss from the measured signal enhancement, was −8.8 dB before annealing and −4.1 dB after annealing.

Higher signal enhancement and net gain can be achieved by optimizing the device design, Er distribution, and post-implantation processing conditions. The present performance, corresponding to net gains of −8.8 dB before annealing and −4.1 dB after annealing, is primarily limited by the overlap between the optical mode and the Er ions, which is reduced by the considerable separation of approximately 80 nm between the centre of the Er ion distribution and the optical mode core. This overlap could be improved by increasing the implantation depth toward the $Si_3N_4$ waveguide, while avoiding structural damage to the waveguide (not withstanding the benefits of limiting the current implantation energy to 200keV). Increasing the $TeO_2$ film thickness could also enhance optical confinement in the Er-doped $TeO_2$ region by shifting the optical mode closer to the surface. In addition, higher Er concentrations are expected to improve signal enhancement, provided that concentration-related quenching is carefully managed. This could be achieved by increasing the implantation dose or by using successive implants with incrementally higher energies to produce a more homogeneous Er distribution over a broader depth. Further reduction of the background propagation loss may be achieved through systematic optimization of the annealing duration and atmosphere. Furthermore, pumping at 980 nm can theoretically provide higher population inversion. The overall device performance also depends on pump and signal mode confinement, coupling efficiency, mode overlap, and propagation loss. Minimizing sidewall interaction remains important for reducing scattering loss. Therefore, the device design must balance implantation parameters, waveguide geometry, Er population dynamics, optical confinement, and propagation loss. Finally, Er/Yb co-implantation could provide an additional route to significantly enhance Er efficiency through efficient energy transfer from Yb to Er ions[28–30].

## 5. Conclusion

In conclusion, this study demonstrates Er implantation into a $TeO_2$ thin film integrated on a silicon nitride photonic platform, followed by low-temperature annealing for post-implantation recovery. Importantly, the study is a first demonstration of an active ion implanted tellurite waveguide to our knowledge. The results show that annealing at 150 °C can partially repair implantation-induced defects, reduce background propagation loss, significantly increase the $Er^{3+}$ photoluminescence lifetime, and improve signal enhancement. These findings highlight the feasibility of using ion implantation combined with low-temperature annealing to realize rare-earth-doped $TeO_2$ thin films compatible with back-end photonic integration. The observed signal enhancement provides an important first step toward integrated tellurite optical amplifiers based on implanted rare-earth ions. Further optimization of implantation depth, Er concentration, ion distribution, and annealing conditions and co-doping with other active dopants could enable higher signal enhancement and net gain in future devices.

## Acknowledgements

The authors acknowledge the support of the Natural Sciences and Engineering Research Council of Canada and Canadian Foundation for Innovation. The authors would like to thank the Centre for Emerging Device Technologies at McMaster University and Doris Stevanovic for support in the thin film depositions, LioniX International for assistance with layout and fabrication, and Dr. Luke Antwis at the UK Ion Beam Centre at the University of Surrey for facilitating the ion implantation. Authors Y.G. and B.H. contributed equally to this work.

## References

[1] Mears R J, Reekie L, Jauncey I M and Payne D N 1987 Low-noise erbium-doped fibre amplifier operating at 1.54μm *Electron. Lett. (UK)* **23** 1026–8

[2] Soares O D D and Anacleto J M S 2000 Erbium Doped Fiber Amplifers Fundamental and Technology *Optics & Laser Technology* **32** 213–4

[3] Nykolak G, Haner M, Becker P C, Shmulovich J and Wong Y H 1993 Systems evaluation of an Er3+ doped planar waveguide amplifier *IEEE Photon. Technol. Lett.* **5** 1185–7

[4] Kenyon A J 2005 Erbium in silicon *Semicond. Sci. Technol.* **20** R65–84

[5] Chen Z, Wan L, Gao S, Zhu K, Zhang M, Li Y, Huang X and Li Z 2022 On-Chip Waveguide Amplifiers for Multi-Band Optical Communications: A Review and Challenge *J. Lightwave Technol.* **40** 3364–73

[6] Shekhar S, Bogaerts W, Chrostowski L, Bowers J E, Hochberg M, Soref R and Shastri B J 2024 Roadmapping the next generation of silicon photonics *Nat Commun* **15** 751

[7] He X, Zhang Z, Ma D, Zhou C, Hou H, Shuai Y, Liu J, Wang R, Zhou Z and Chen W 2025 Erbium-doped/erbium-ytterbium co-doped waveguide amplifiers in silicon-based optoelectronics: recent progress *Adv. Photon.* **7**

[8] Wang S, Yang L, Cheng R, Xu Y, Shen M, Cone R L, Thiel C W and Tang H X 2020 Incorporation of erbium ions into thin-film lithium niobate integrated photonics *Applied Physics Letters* **116** 151103

[9] Liang Y, Zhou J, Liu Z, Zhang H, Fang Z, Zhou Y, Yin D, Lin J, Yu J, Wu R, Wang M and Cheng Y 2022 A high-gain cladded waveguide amplifier on erbium doped thin-film lithium niobate fabricated using photolithography assisted chemo-mechanical etching *Nanophotonics* **11** 1033–40

[10] Bao R, Fang Z, Liu J, Liu Z, Chen J, Wang M, Wu R, Zhang H and Cheng Y 2025 An Erbium-Doped Waveguide Amplifier on Thin Film Lithium Niobate with an Output Power Exceeding 100 mW *Laser & Photonics Reviews* **19** 2400765

[11] Zhang H, He Y, Zhu S, Zhou X and Zhang L 2025 Atomic-Layer Engineered Erbium-Doped Waveguide Amplifier with a 14.4 dB Net Gain *ACS Photonics* **12** 674–83

[12] Cai M, Wu K, Xiang J, Xiao Z, Li T, Li C and Chen J 2022 Erbium-Doped Lithium Niobate Thin Film Waveguide Amplifier With 16 dB Internal Net Gain *IEEE J. Select. Topics Quantum Electron.* **28** 1–8

[13] Cai M, Li T, Zhang X, Zhang H, Wang L, Li H, Zheng Y, Chen X, Chen J and Wu K 2024 Gain Dynamics in Integrated Waveguide Amplifier Based on Erbium-Doped Thin-Film Lithium Niobate *ACS Photonics* **11** 4923–32

[14] Han J, Li M, Wu R, Yu J, Gao L, Fang Z, Wang M, Liang Y, Zhang H and Cheng Y 2025 High fiber-to-fiber net gain in erbium-doped thin film lithium niobate waveguide amplifier as an external gain chip *Opto-Electron Sci* **4** 250004

[15] Wang Y, Shen B, Wang B, Yang S, Yao L, Chen R, Zhang Y, Wang H, Zhang X, Zhou P, Tao Z, Xing L, Lin Z, Wu Y, Li W, Sun D, Shu H and Wang X 2025 Unifying optical gain and electro-optical dynamics in Er-doped thin-film lithium niobate platform *Nat Commun* **16** 10462

[16] Rönn J, Zhang W, Autere A, Leroux X, Pakarinen L, Alonso-Ramos C, Säynätjoki A, Lipsanen H, Vivien L, Cassan E and Sun Z 2019 Ultra-high on-chip optical gain in erbium-based hybrid slot waveguides *Nat Commun* **10** 432

[17] Demirtas M and Ay F 2020 High-Gain $Er^{3+}$:$Al_2O_3$ On-Chip Waveguide Amplifiers *IEEE J. Select. Topics Quantum Electron.* **26** 1–8

[18] Bonneville D B, Osornio-Martinez C E, Dijkstra M and García-Blanco S M 2024 High on-chip gain spiral $Al_2O_3$:$Er^{3+}$ waveguide amplifiers *Opt. Express* **32** 15527

[19] Osornio-Martinez C E, Bonneville D B, Dijkstra M, Do Nascimento Jr. A R and García-Blanco S M 2025 Broadband Packaged Erbium-Doped Polycrystalline $Al_2O_3$ Waveguide Amplifier with 24 dB External Net Gain *Opt. Express* **33** 28985

[20] Osornio-Martinez C E, Bonneville D B, Hegeman I, Dijkstra M, Segondat Q, Dekker R and García-Blanco S M 2025 Monolithically integrated erbium-doped polycrystalline $Al_2O_3$ waveguide amplifier on silicon photonics platform *Opt. Express* **33** 23491

[21] Osornio-Martinez C E, Bonneville D B, Dijkstra M and García-Blanco S M 2025 Scalable erbium-doped waveguide amplifier with external fiber-to-fiber gain using reactively sputtered polycrystalline aluminium oxide *Opt. Express* **33** 22458

[22] Mu J, Dijkstra M, Korterik J, Offerhaus H and García-Blanco S M 2020 High-gain waveguide amplifiers in $Si_3N_4$ technology via double-layer monolithic integration *Photon. Res.* **8** 1634

[23] Vázquez-Córdova S A, Dijkstra M, Bernhardi E H, Ay F, Wörhoff K, Herek J L, García-Blanco S M and Pollnau M 2014 Erbium-doped spiral amplifiers with 20 dB of net gain on silicon *Opt. Express* **22** 25993

[24] Subramanian A Z, Murugan G S, Zervas M N and Wilkinson J S 2012 High index contrast Er:Ta2O5 waveguide amplifier on oxidised silicon *Optics Communications* **285** 124–7

[25] Zhang Z, Liu R, Wang W, Yan K, Yang Z, Song M, Wu D, Xu P, Wang X and Wang R 2023 On-chip Er-doped $Ta_2O_5$ waveguide amplifiers with a high internal net gain *Opt. Lett.* **48** 5799

[26] Sun Y, Cheng Q, Xia L, Wang Z, Chen P, Long Z, Wang R and Zou Y 2026 Large-mode area erbium-doped tantalum pentoxide waveguide amplifier with high on-chip net gain *Opt. Lett.* **51** 452

[27] Shui L, Yi R, Zhao C, Lu J, Zhang X, Zhang J and Gan X 2026 On-Chip Erbium-Doped Tantalum Oxide Microring Hybrid Cavity Single-Mode Laser

[28] Guo R, Wang X, Zang K, Wang B, Wang L, Gao L and Zhou Z 2011 Optical amplification in Er/Yb silicate strip loaded waveguide *Applied Physics Letters* **99** 161115

[29] Lei Wang, Ruimin Guo, Bing Wang, Xingjun Wang, and Zhiping Zhou 2012 Hybrid Si3N4-Er/Yb Silicate Waveguides for Amplifier Application *IEEE Photon. Technol. Lett.* **24** 900–2

[30] Guo R, Wang B, Wang X, Wang L, Jiang L and Zhou Z 2012 Optical amplification in Er/Yb silicate slot waveguide *Opt. Lett.* **37** 1427

[31] Sun H, Yin L, Liu Z, Zheng Y, Fan F, Zhao S, Feng X, Li Y and Ning C Z 2017 Giant optical gain in a single-crystal erbium chloride silicate nanowire *Nature Photon* **11** 589–93

[32] Vu K and Madden S 2010 Tellurium dioxide Erbium doped planar rib waveguide amplifiers with net gain and 28dB/cm internal gain *Opt. Express* **18** 19192

[33] Vu K, Farahani S and Madden S 2015 980nm pumped erbium doped tellurium oxide planar rib waveguide laser and amplifier with gain in S, C and L band *Opt. Express* **23** 747

[34] Frankis H C, Mbonde H M, Bonneville D B, Zhang C, Mateman R, Leinse A and Bradley J D B 2020 Erbium-doped $TeO_2$ -coated $Si_3N_4$ waveguide amplifiers with 5 dB net gain *Photon. Res.* **8** 127

[35] Mbonde H M, Hashemi B, Segat Frare B L, Wildi T, Ahmadi P T, Bonneville D B, Singh N, Mascher P, Kärtner F X, Herr T and Bradley J D B 2025 Demonstration of passive, nonlinear, and active devices on a hybrid photonic platform *Opt. Express* **33** 1836

[36] Guo Y, Jiang W, Bao X, Zhou R, Guo L, Zhang B, Chen W and Zhou L 2026 Photoluminescence enhancement in Er- and Er/Yb-implanted silicon nitride waveguides via process-property engineering *Opt. Express* **34** 30472

[37] Polman A, Jacobson D C, Eaglesham D J, Kistler R C and Poate J M 1991 Optical doping of waveguide materials by MeV Er implantation *Journal of Applied Physics* **70** 3778–84

[38] Van Den Hoven G N, Snoeks E, Polman A, Van Dam C, Van Uffelen J W M and Smit M K 1996 Upconversion in Er-implanted Al2O3 waveguides *Journal of Applied Physics* **79** 1258–66

[39] Polman A 1997 Erbium implanted thin film photonic materials *Journal of Applied Physics* **82** 1–39

[40] Liu Y, Qiu Z, Ji X, Lukashchuk A, He J, Riemensberger J, Hafermann M, Wang R N, Liu J, Ronning C and Kippenberg T J 2022 A photonic integrated circuit–based erbium-doped amplifier *Science* **376** 1309–13

[41] Qiu Z, Ji X, Liu Y, Hafermann M, Kim T, Olson J C, Ning W R, Ronning C and Kippenberg T 2024 Hybrid integrated multi-lane erbium-doped Si3N4 waveguide amplifiers *Optical Fiber Communication Conference (OFC) 2024* Optical Fiber Communication Conference (San Diego California: Optica Publishing Group) p M4A.5

[42] Liu Y 2024 Erbium-Doped $Si_3 N_4$ Photonic Integrated Circuits and Wafer-Scale Fabrication to Include our Recent Progress *Optical Fiber Communication Conference (OFC) 2024* Optical Fiber Communication Conference (San Diego California: Optica Publishing Group) p Th1D.1

[43] Slusar T, Azarov A, Galeckas A, Hallén A, Ju J J, Moon K and Kuznetsov A 2024 Er:$LiNbO_3$ Quantum Memory Platform Optimized with Dynamic Defect Annealing *Advanced Optical Materials* **12** 2401374

[44] Wang L, Lerner S, Zhu H, Cui X, Zeng B, Wei F, Wendler E and Ronning C 2025 Intense Photoluminescence in Erbium-Ion-Implanted Lithium Niobate Thin Films and Its Interplay with Lattice Defects *Advanced Optical Materials* **13** e02568

[45] Dearnaley G 1975 Ion implantation *Nature* **256** 701–5

[46] Qiu Z, Yang X, Li X, Hu J, Liu Z, Zhang Y, Ji X, Sun J, Lihachev G, Li Z, Kentsch U and Kippenberg T J 2026 High-pulse-energy integrated mode-locked laser using a Mamyshev oscillator *Nature* **654** 57–63

[47] Ji X, Yang X, Liu Y, Qiu Z, Lihachev G, Bianconi S, Sun J, Voloshin A, Kim T, Olson J C and Kippenberg T J 2026 Wafer-scale manufacturing of ultra-broadband, high-power erbium-doped integrated lasers *Nat Commun* **17** 3722

[48] Klaver Y, Te Morsche R, Botter R A, Hashemi B, Segat Frare B L, Mishra A, Ye K, Mbonde H M, Torab Ahmadi P, Majidian Taleghani N, Jonker E, Braamhaar R B G, Selvaganapathy P R, Mascher P, Van Der Slot P J M, Bradley J D B and Marpaung D 2026 Surface acoustic wave Brillouin photonics on a silicon nitride chip *Nat. Photon.* **20** 637–43

[49] Frankis H C, Kiani K M, Su D, Mateman R, Leinse A and Bradley J D B 2019 High-Q tellurium-oxide-coated silicon nitride microring resonators *Opt. Lett.* **44** 118

[50] Segat Frare B L, Torab Ahmadi P, Hashemi B, Bonneville D B, Mbonde H M, Frankis H C, Knights A P, Mascher P and Bradley J D B 2023 On-chip hybrid erbium-doped tellurium oxide–silicon nitride distributed Bragg reflector lasers *Appl. Phys. B* **129** 158

[51] Segat Frare B L, Hashemi B, Majidian Taleghani N, Torab Ahmadi P, Bonneville D B, Mbonde H M, Frankis H C, Mascher P, Selvaganapathy P R and Bradley J D B 2025 Thulium-doped tellurite distributed Bragg reflector waveguide laser on a silicon nitride chip *J. Opt. Soc. Am. B* **42** 1204

[52] Blumenthal D J, Heideman R, Geuzebroek D, Leinse A and Roeloffzen C 2018 Silicon Nitride in Silicon Photonics *Proc. IEEE* **106** 2209–31

[53] Xiang C, Jin W and Bowers J E 2022 Silicon nitride passive and active photonic integrated circuits: trends and prospects *Photon. Res.* **10** A82

[54] Buzaverov K A, Baburin A S, Sergeev E V, Avdeev S S, Lotkov E S, Bukatin S V, Stepanov I A, Kramarenko A B, Amiraslanov A Sh, Kushnev D V, Ryzhikov I A and Rodionov I A 2024 Silicon Nitride Integrated Photonics from Visible to Mid-Infrared Spectra *Laser & Photonics Reviews* 2400508

[55] Singh N, Mbonde H M, Frankis H C, Ippen E, Bradley J D B and Kärtner F X 2020 Nonlinear silicon photonics on CMOS-compatible tellurium oxide *Photon. Res.* **8** 1904

[56] Bu M-Y, Zhao J-H, Wang S-M, Gu J-J, Fan L, Ren Y-Y and Zhang L 2022 Optical properties investigation on Er3+ ion-doped tellurite glass by hydrogen and oxygen ion implantation *Journal of Luminescence* **248** 118882

[57] Berneschi S, Nunzi Conti G, Bányász I, Watterich A, Khanh N Q, Fried M, Pászti F, Brenci M, Pelli S and Righini G C 2007 Ion beam irradiated channel waveguides in Er3+-doped tellurite glass *Applied Physics Letters* **90** 121136

[58] Liu C-X, Xu J, Zhou Z-G, Fu L-L, Zheng R-L, Li W-N, Guo H-T, Lin S-B and Wei W 2015 Planar waveguides in $Yb^{3+}$ -doped tellurite glasses fabricated by ion implantation of oxygen *Mod. Phys. Lett. B* **29** 1550021

[59] Roeloffzen C G H, Hoekman M, Klein E J, Wevers L S, Timens R B, Marchenko D, Geskus D, Dekker R, Alippi A, Grootjans R, Van Rees A, Oldenbeuving R M, Epping J P, Heideman R G, Worhoff K, Leinse A, Geuzebroek D, Schreuder E, Van Dijk P W L, Visscher I, Taddei C, Fan Y, Taballione C, Liu Y, Marpaung D, Zhuang L, Benelajla M and Boller K-J 2018 Low-Loss Si3N4 TriPleX Optical Waveguides: Technology and Applications Overview *IEEE J. Select. Topics Quantum Electron.* **24** 1–21

[60] Preston K, Schmidt B and Lipson M 2007 Polysilicon photonic resonators for large-scale 3D integration of optical networks *Opt. Express* **15** 17283

[61] Dewan N, Sreenivas K and Gupta V 2007 Properties of crystalline γ-TeO2 thin film *Journal of Crystal Growth* **305** 237–41

[62] Siciliano T, Di Giulio M, Tepore M, Filippo E, Micocci G and Tepore A 2010 Effect of thermal annealing time on optical and structural properties of TeO2 thin films *Vacuum* **84** 935–9

[63] Kumar S and Mansingh A 1990 Annealing-induced structural changes in tellurium dioxide thin films *J. Phys. D: Appl. Phys.* **23** 1252–5

[64] Mahendar M, Krishna R N V and Chaudhary A K 2023 Effect of the annealing temperature on structural, morphological, and nonlinear optical properties of $TeO_2$ thin films used for efficient THz generation *Appl. Opt.* **62** 2394

[65] Li J, Hu H and Gan F 2003 Influence of OH groups on 1.5-μm emission of $Yb^{3+}$/$Er^{3+}$ co-doped tungsten-tellurite glasses *Chinese Optics Letters* **1**

[66] Dai S, Zhang J, Yu C, Zhou G, Wang G and Hu L 2005 Effect of hydroxyl groups on nonradiative decay of Er3+:4I13/2→4I15/2 transition in zinc tellurite glasses *Materials Letters* **59** 2333–6

[67] Coffa S, Priolo F and Battaglia A 1993 Defect production and annealing in ion-implanted amorphous silicon *Phys. Rev. Lett.* **70** 3756–9